\documentclass[manuscript]{aastex}

\usepackage{color}
\usepackage{graphicx}
\usepackage{graphicx,amsmath}

\shorttitle{Orbits and Masses of the Inner Satellites of Saturn}
\shortauthors{Cooper et al.}

\begin{document}

\title{Saturn's inner satellites : orbits, masses and the chaotic motion of Atlas from new {\it Cassini} imaging observations }

\author{N.J. Cooper\altaffilmark{1,2,3}, S. Renner\altaffilmark{2,3}, C.D.Murray\altaffilmark{1}, 
M.W. Evans\altaffilmark{4}}

\altaffiltext{1}{Astronomy Unit, School of Physics and Astronomy, Queen Mary University of London, 
Mile End Road, London, E1 4NS, UK}
\altaffiltext{2}{Universit\'e Lille 1, Laboratoire d'Astronomie de Lille (LAL), 
1 impasse de l'Observatoire, 59000 Lille, France}
\altaffiltext{3}{Institut de M\'ecanique C\'eleste et de Calcul des Eph\'em\'erides (IMCCE), 
Observatoire de Paris, UMR 8028 du CNRS, 77 avenue Denfert-Rochereau, 75014 Paris, France}
\altaffiltext{4}{Department of Astronomy, Cornell University, Ithaca, NY 14853, USA}

\slugcomment{Accepted for publication in The Astronomical Journal}

\catcode`±=\active
\def±{\phantom{$-$}}
\catcode`?=\active
\def?{\phantom{0}}

\begin{abstract}
We present numerically-derived orbits and mass estimates for the inner Saturnian satellites, Atlas, Prometheus, Pandora, Janus and Epimetheus from a fit to 2580 new Cassini ISS astrometric observations spanning February 2004 to August 2013. The observations are provided in a supplementary table. We estimate $GM_{\rm Atlas}$$=$(0.384$\pm$0.001$)\times$10$^{-3}$km$^3$ s$^{-2}$, a value 13$\%$ smaller than the previously published estimate but with an order of magnitude reduction in the uncertainty. We also find $GM_{\rm Prometheus}$$=$(10.677$\pm$0.006$)\times$10$^{-3}$km$^3$ s$^{-2}$, $GM_{\rm Pandora}$$=$(9.133$\pm$0.009$)\times$10$^{-3}$km$^3$ s$^{-2}$, $GM_{\rm Janus}$$=$(126.51$\pm$0.03$)\times$10$^{-3}$km$^3$ s$^{-2}$ and $GM_{\rm Epimetheus}$$=$(35.110$\pm$0.009$)\times$10$^{-3}$km$^3$ s$^{-2}$, consistent with previously published values, but also with significant reductions in uncertainties. We show that Atlas is currently librating in both the 54:53 co-rotation-eccentricity resonance (CER) and the 54:53 inner Lindblad (ILR) resonance with Prometheus, making it the latest example of a coupled CER-ILR system, in common with the Saturnian satellites Anthe, Aegaeon and Methone, and possibly Neptune's ring arcs. We further demonstrate that Atlas's orbit is chaotic, with a Lyapunov time of $\sim$10 years, and show that its chaotic behaviour is a direct consequence of the coupled resonant interaction with Prometheus, rather than being an indirect effect of the known chaotic interaction between Prometheus and Pandora. We provide an updated analysis of the second-order resonant perturbations involving Prometheus, Pandora and Epimetheus based on the new observations, showing that these resonant arguments are librating only when Epimetheus is the innermost of the co-orbital pair, Janus and Epimetheus. We also find evidence that the known chaotic changes in the orbits of Prometheus and Pandora are not confined to times of apse anti-alignement.

\end{abstract}


\clearpage

\section{Introduction}

The complexity of the dynamical environment occupied by the small inner Saturnian satellites is now becoming clearer, thanks in large part to the success of observing campaigns such as those using the Imaging Science Subsystem (ISS) of the $Cassini$ orbiter. An understanding of this dynamical environment and its past history is crucial in constraining models for the physical origin and evolution of the inner satellites and rings of Saturn \citep{Lain09, Char11, Lain12}. The availability of high-resolution ISS images to generate a dense coverage of high-quality astrometric observations of these satellites, so far spanning almost 10 years has been vital, while similar campaigns using the Hubble Space Telescope (HST) have provided a temporal link with earlier Voyager observations \citep{French03}. Although in this work, we consider only the current state of five of the inner satellites of Saturn and their short-term dynamical evolution, we also provide high-precision starting conditions and mass estimates which can be used to constrain studies of the longer-term dynamical and physical evolution of the system as a whole.

This work focuses on Atlas, Prometheus, Pandora, Janus and Epimetheus, the five satellites closest to the main rings of Saturn, excluding Daphnis and Pan which orbit within gaps inside the A Ring. We treat these satellites as a group because of their close dynamical relationship, described in more detail below. In previous work, \citet{Spit06} fitted observations of Atlas using Voyager, Hubble Space Telescope (HST) and $Cassini$ ISS observations spanning 2004 FEB 06 and 2005 NOV 06, in addition to Earth-based, Voyager, HST and $Cassini$ observations of Prometheus, Pandora, Janus and Epimetheus. \citet{Jac08} updated the work of \citet{Spit06} with the addition of new $Cassini$ observations, extending the observation time-span to 2007 MAR 24. The extension of the timespan of $Cassini$ ISS observations to 2013 AUG 28 in the current work is particularly important in terms of covering the most recent chaotic interaction between Prometheus and Pandora in February 2013, as well as the latest switch in the orbits of the co-orbitals, Janus and Epimetheus, in January 2010. As indicated below, the additional need to constrain resonant effects in the orbit of Atlas with periods of three or more years also greatly benefits from this extended dense coverage of observations.

Resonant phenomena are widespread within the Saturnian satellite and ring system as a whole, but particularly so amongst the small inner satellites, and a clear picture of the nature and evolution of these resonances is important in unravelling the dynamical history of the inner satellites and rings.  Therefore, in addition to providing updated orbits and mass estimates for these satellites based on the extended timespan provided by the new observations, a further objective of this paper is to provide an updated and more comprehensive numerical analysis of the key resonant relationships between the five satellites forming the subject of this study.

In terms of known resonant relationships, the co-orbitals Janus and Epimetheus undergo a switch in their horseshoe orbital configuration every four years, with the outer satellite of the pair becoming the inner and vice versa. Prometheus and Pandora undergo a chaotic interaction every 6.2 years at closest approach (apse-antialignment), as a result of a 121:118 resonance between these two satellites \citep{Gold03a,Gold03b,Ren03,Coop04,Ren05}. \citet{Farm06}, who further studied the Prometheus$-$Pandora system analytically and numerically, showed that these interactions may occur at times other than at apse-antialignments. Pandora is in a 3:2 near-resonance with Mimas, which causes an oscillation in its longitude with period 612 days, modulated by the approximate 70 year period due to the inclination-type resonance between Mimas and Tethys \citep{French03}. \citet{Spit06} noted a periodic perturbation in the orbit of Atlas with an amplitude of about 600 km and a period of about 3 years, due to a 54:53 mean motion resonance with Prometheus, suggesting that the likely librating resonant argument is 53$\lambda_{\rm Prom}$$-$52$\lambda_{\rm Atlas}$$-$$\varpi_{\rm Prom}$. They also identified a 70:67 mean motion resonance between Pandora and Atlas with an amplitude of about 150 km and argued that, because the orbits of Prometheus and Pandora are chaotic, Atlas is also likely to be chaotic. \citet{Coop04} noted the proximity of Prometheus to the 17:15 mean motion resonance with Epimetheus and Pandora to the 21:19 mean motion resonance, also with Epimetheus, and further noted that these resonances sweep across their orbits when Epimetheus switches position with respect to Janus every four years.

In this paper, we give an updated analysis of these resonant relationships, based on the new orbital solutions. For example, we show not only that Atlas is currently librating in the 54:53 mean motion resonance with Prometheus with resonant argument 53$\lambda_{\rm Prom}$$-$52$\lambda_{\rm Atlas}$$-$$\varpi_{\rm Prom}$, but that in addition, the resonant argument 53$\lambda_{\rm Prom}$$-$52$\lambda_{\rm Atlas}$$-$$\varpi_{\rm Atlas}$ is also librating. Thus we find that Atlas can be added to a growing list of small inner Saturnian satellites that are simultaneously librating in coupled corotation-eccentricity (CER) and inner Lindblad (ILR)-type resonances. The other examples so far known are Methone, Anthe and Aegaeon interacting with Mimas \citep{Spit06, Coop08, Hed09,Elmout14} and possibly Neptune's ring arcs \citep{Nam02,Ren14}. We then show that the observed short-term dynamical evolution of Atlas can be well-approximated by a three-body system consisting of Atlas, Prometheus and Saturn only, and use a chaos indicator to estimate the chaoticity in Atlas' orbit. Hence the coupled first-order resonant interaction between Atlas and Prometheus is the primary dynamical influence on the orbit of Atlas. It follows that the primary source of chaos in the orbit of Atlas is the direct effect of the coupled resonance with Prometheus, rather than the secondary effect of the chaotic interaction between Prometheus and Pandora. 

\section{Observations}

Cassini ISS images used in this work come mainly from SATELLORB image sequences. These sequences form part of a campaign of observations of the small inner satellites of Saturn in progress throughout the Cassini tour, a campaign which by end-of-mission will have provided dense coverage between 2004 and 2017. In addition, we include observations from image sequences designed to study Saturn's F ring, many of which contain opportunistic sightings of Atlas, Prometheus and Pandora, in addition to Pan and Daphnis (though these satellites do not form part of the current work). Of the 2580 observations presented here, the majority (2567) used narrow angle camera (NAC) images while the remaining 13 came from wide-angle camera (WAC) images.

Astrometric measurements were made using the Caviar software package. Caviar (CAssini Visual Image Analysis Release) was developed at Queen Mary University of London for the analysis and astrometric reduction of Cassini ISS images and uses the IDL data language (Exelis Visual Information Solutions, Boulder, Colorado). Data reduction consisted of a correction to the camera pointing orientation together with a measurement of the pixel coordinates (line and sample) of the astrometric position of each satellite in a given image (we use the Cassini convention of referring to the x-coordinate as 'sample' and y-coordinate as 'line'). The camera pointing was corrected using the positions of reference stars extracted from the UCAC2 and Tycho-2 catalogues \citep{Zach04,hog00}. We obtain a typical pointing accuracy of 0.1 pixel (0.12357 arc sec) or less. Zacharias et al (2013) quote positional differences between UCAC2 and the more recent UCAC4 catalogue of Ôonly a few up to 10 mas over the entire rangeÕ of magnitude, so we would not expect our achieved pointing accuracy to change significantly with UCAC4. The astrometric positions of unresolved or poorly resolved satellites were estimated using a centroid-estimation method based on \citet{Stet87}. The measured positions of the centre-of-light were then adjusted to the centre-of-figure using a correction depending on the phase angle between the observer, satellite and the Sun. For images of resolved satellites, the centre-of-figure was estimated by fitting an ellipsoidal shape model obtained from the latest SPICE kernels (Table 7). The number of limb-fitted and centroided measurements for each satellite are summarised in Table 1. 

The complete set of observations is provided as a machine-readable table, published in the electronic edition of this paper. In this table, we provide for each observation, the image name, time, estimated pointing information in terms of the right ascension, declination and TWIST angles for of the optical axis, the measured line and sample values for the satellite in question, and their equivalents expressed as right ascension and declination. A small section of the full table is provided in Table 2 as an indication of its form and content.

\section{Orbital Solutions}

A numerical integration of the full equations of motion in 3D was fitted to the astrometric data for Atlas, Prometheus, Pandora, Janus and Epimetheus, solving for their state vectors at the epoch 2007 JUN 01 00:00:00.0, and for their masses. The perturbing effects of Mimas, Enceladus, Tethys, Dione, Rhea, Titan, Hyperion and Iapetus were included in the model, using positions extracted from JPL's SAT351 ephemeris, via the SPICE package \citep{Act96}, together with the mass values given in Table 4. Saturn's oblateness was incorporated up to terms in J6 (values also from SAT351), while perturbations from the Jupiter system and the Sun were based on positions from DE414 (Table 3). 

Solutions for the state vectors and masses were obtained by solving the variational equations simultaneously with the equations of motion, minimising the observed-minus-computed (O-C) residuals iteratively. All observations were equally weighted. Numerical integration of the equations of motion and the variational equations was performed using the 12th-order Runge-Kutta-Nystr\"om RKN12(10)17M algorithm of \citet{brank87}, while the equations of condition were solved using the SVD-based approach of Lawson and Hanson \citep{Lh75}.  For more details relating to the numerical integration and model fitting scheme used in this work, see \citet{Mur05}. The final solutions for the fitted state vectors are given in Table 5 and for the masses (GM values) in Table 6.

\subsection{Analysis of Residuals}

Figures 1 and 2, respectively, compare the pre- and post-fit O-C residuals, in pixels, for each satellite, plotted as a function of time. The pre-fit residuals (Figure 1) were computed with respect to the JPL ephemeris SAT353, while the post-fit residuals (Figure 2) are those obtained after fitting the numerical model described in the previous section to the observations. We compare with SAT353 rather than the recently-released (January 2014) SAT363 epheremis in order to be able to show a pre/post-fit comparison. SAT363 is based on a fit to Cassini data up to late 2013, including the SATELLORB observations presented in this paper, so residuals relative to this ephemeris are post-fit.

The improvement in the post-fit residuals (Figure 2) for the later observations compared their pre-fit equivalent (Figure 1) is to be expected because the SAT353 ephemeris is based on a fit to Earth-based, Voyager, HST and Cassini data up to 2011 JAN 20 and is therefore unable to constrain adequately the chaotic interaction between Prometheus and Pandora which occurred around 2013 FEB 18. The size of the pre-fit residuals for the later observations is a direct result of the chaotic orbital motion of Atlas, Prometheus and Pandora, and underlines the need for good observational coverage for these satellites. 

The RMS O-C residual values for each satellite in the line and sample directions, are provided in Table 1, in units of km. The equivalent mean values, in units of pixels, are labelled in Figures 1 and 2 and indicate the presence of a small positive bias in the line residuals for Prometheus, Pandora, Janus and Epimetheus, though not for Atlas. Mean sample residuals are close to zero, indicating no significant bias. This is also evident in Figure 3, where the residuals are plotted as line versus sample, converted to km. Investigation has shown that the origin of this bias is in the use of limb-detection to find the centre-of-figure from resolved satellite images. The position of the centre-of-figure was found to be slightly biased in the direction of the satellite limb, and this was found to be true to a greater-or-lesser extent with three different limb detection techniques. If the satellite limb directions were randomly distributed from image to image, no systematic effect would be detectable. However because limbs are dominantly aligned in the same direction in the majority of images, for spacecraft operational reasons, this appears as a systematic effect in the positive line direction \citep{Coop14}. Atlas observations were mainly centroided rather than limb-fitted (Table 1) and thus mean residuals for Atlas are close to zero. Again, comparing Figures 1 and 2, the mean values for Prometheus, Pandora, Janus and Epimetheus do not change significantly, post-fit. This suggests that any bias in the observations is not being absorbed by the fit and therefore not significantly affecting the results. This was confirmed by applying a pre-fit correction to the observations based on the average of the pre/post-fit mean line residual values for each of the satellites Prometheus, Pandora, Janus and Epimetheus (0.39, 0.27, 0.64 and 0.45 pixels respectively), before rerunning the fit. The fitted masses obtained for all the satellites, including Atlas, changed only in the last decimal place, and the changes were, in all cases, within the stated parameter uncertainties.

\section{Numerical Modelling}

Using the full numerical model derived from the fit to the observations, we generated ephemerides for Atlas, Prometheus, Pandora, Janus and Epimetheus spanning the period 2000 to 2020. Given that the observation timespan is approximately 2004 to 2013, the ephemerides include a period of a few years at each end of the time-range which lack observational constraint. However, we believe that this is justifiable given the precision of the observations and that the Lyapunov time for Atlas-Prometheus system is approximately 10 years (this work, see below) and for the Prometheus-Pandora system is approximately 3 years (this work and \citet{Gold03a,Coop04}).

\subsection{Orbital Elements}

In Figures 4 to 8 inclusive, we show the geometrical orbital elements for Atlas, Prometheus, Pandora, Janus and Epimetheus respectively, plotted as a function of time, between 2000 JAN 1 and 2020 JAN 1. These were derived by converting state vectors, extracted from the ephemeris at 1.5 hour intervals, to geometric elements using the algorithm of \citet{Ren06}. 

In Figure 4, we note an irregular periodic oscillation in the semi-major axis, eccentricity, mean longitude and longitude of pericentre of Atlas. We show below that this is the effect of a couple CER-ILR resonant relationship between Atlas and Prometheus. We also show that the irregularity of the oscillation is an effect of the chaotic nature of this resonant interaction.

The orbit of Prometheus is known to be chaotic and in Figure 5, the irregularity of Prometheus's orbital elements as a function of time is immediately apparent, with no well-behaved periodic effects visible on this twenty-year timescale. By comparison, the periodic effect of the near-resonance between Pandora and Mimas, described in the introduction, is clear in Figure 6 and this tends to mask the expected anti-correlation between the mean longitudes of Prometheus and Pandora (compare Figures 5(d) and 6(d).

Comparing Figures 7 and 8, the orbital elements for the co-orbitals Janus and Epimetheus are dominated by the effects of the four-yearly switch in their orbital configuration. The sharp, linear saw-tooth behaviour of the mean longitudes (Figures 7(d) and 8(d)) suggests that the mean motions of these satellites are constant in between otherwise virtually instantaneous switches in their orbital configuration. However this is misleading:  the true behaviour is more subtle and cannot be seen in the mean longitudes on this scale. The semi-major axes (Figures 7(a) and 8(a)) provide a more revealing picture of the variability of the orbital motion, clearly showing that the semi-major axes and hence the mean motions are in fact constantly changing over the course each four-year cycle. Thus the orbits of these satellites do not behave as uniformly precessing ellipses in between each switch in configuration. See also Figure 3.25, p.144 from \citet{md99}.

\subsection{Resonant Perturbations}

\citet{French03,Gold03a,Coop04,Spit06} independently inferred the existence of a variety of mean-motion resonant relationships between Atlas, Prometheus, Pandora and Epimetheus based on fixed estimates of their semi-major axes, apsidal and nodal precession rates. In practice, these three quantities all vary with time and the only complete way to confirm whether or not a state of resonance exists at any time is to plot the time-variation of the resonant argument in question. Below, we describe each of these resonant relationships in more detail and display each resonant argument as a function of time, using values extracted from the full multi-body 3D numerical model, described earlier.

\subsubsection{Epimetheus, Prometheus and Pandora}
Based on their respective semi-major axes, apsidal and nodal precession rates, \citet{Coop04} identified two sets of second-order mean motion resonances due to Epimetheus close to the orbits of Prometheus and Pandora: specifically, three 17:15 eccentricity-type resonances between Epimetheus and Prometheus and three 21:19 eccentricity-type resonances between Epimetheus and Pandora (\citet{Coop04} Table 5). They also showed that the locations of these resonance change with respect to Prometheus and Pandora every four years, when Janus and Epimetheus undergo their four-yearly switch in co-orbital configuration. However, \citet{Coop04} did not plot the time-variation of the these resonant arguments. We do so in Figure 9, where we see very clearly that in all cases these resonant arguments change their behaviour precisely at times corresponding to the four-year switches in the orbital configuration of Janus and Epimetheus, marked in the figure by the vertical dot-dashed red lines. 

In each of the periods 2002-2006, 2010-2014, 2018-2020, when Epimetheus is the innermost of the two co-orbitals, all six resonant arguments are either in a state of libration or slow-circulation, while in between, when Epimetheus is outermost, all six arguments are rapidly circulating and no state of resonance exists. More specifically, we see that the arguments $17\lambda^{\prime}-15\lambda-\varpi^{\prime}-\varpi$ (where the primed quantity refers to Epimetheus and the unprimed to Pandora) and $21\lambda^{\prime}-19\lambda-2\varpi$ (where the unprimed quantity now refers to Pandora) are both librating when Epimetheus is in the inner configuration. Thus Epimetheus is simultaneously in a second-order mean motion resonance with both Prometheus and Pandora when in the inner configuration, and outside this resonance when in the outer configuration with respect to Janus.

\subsubsection{Prometheus and Pandora}
\citet{Gold03a,Ren03} showed that, approximately every 6.2 years, at closest approach (apse anti-alignment), Prometheus and Pandora interact gravitationally in such a way that their mean motions change chaotically, thus accounting for the observed drift in their mean longitudes identified by \citet{French03} in Hubble Space Telescope (HST) observations, compared to earlier Voyager observations. \citet{Gold03a,Ren03} identified four overlapping 121:118 mean motion resonances as being the source of the chaos in the orbits of Prometheus and Pandora. Using a pendulum model \citet{Farm06} showed analytically that these observed sudden changes in mean motion represent separatrix crossings, marking short-periods of libration followed by longer periods of circulation. However, they further demonstrated numerically, using the full equations of motion, that these separatrix crossings are not in fact confined to times of closest approach at apse anti-alignment, and that they can occur far from such times, even at times of apse-alignment.

This seems to be confirmed both by Figure 10, showing the mean longitudes for Atlas, Prometheus and Pandora, where the sudden jumps in the mean longitude of Prometheus do not always seem to correlate with apse anti-alignments (vertical black dashed lines) and by Figure 11, in which we plot as a function of time the four resonant arguments for the 121:118 mean motion resonance identified by \citet{Gold03a,Ren03}, using values derived from our full numerical model. The times of apse anti-alignments marked by the vertical black dashed lines in Figure 11 also do not correlate in any obvious way with the boundaries between episodes of libration and circulation of the resonant arguments. This apparently confirms the conclusion of \citet{Farm06} that these jumps should not be confined to times of anti-alignment alone.

\subsubsection{Atlas and Prometheus}

Referring again to Figure 10, in which the mean longitudes for Atlas, Prometheus and Pandora are plotted as a function of time in the same display, the vertical black lines mark the times of closest approach between Prometheus and Pandora, approximately every 6.2 years, while the vertical red lines mark the times of the switches in the orbital configuration of janus and Epimetheus, every 4 years. The fit epoch is marked by the blue vertical line. The mean longitude of Atlas is dominated by a $\sim$4-year oscillation, which we show below is the effect of the 54:53 CER with Prometheus. 

Figure 12 displays the temporal variation of the resonant arguments $ \phi_1=54\lambda^{\prime}-53\lambda-\varpi $ and $ \phi_2=54\lambda^{\prime}-53\lambda-\varpi^{\prime} $, again, using the orbital elements derived from the full numerical model as above. Primed quantities refer to Prometheus and unprimed quantities to Atlas, with $\lambda$ representing the mean longitude, and $\varpi$ the longitude of pericentre in each case. Once again, vertical black dashed lines mark closest approaches between Prometheus and Pandora, vertical red dot-dashed lines mark switches in the configuration of Janus and Epimetheus and the vertical blue line marks the fit epoch. We see that both arguments are librating in an irregular fashion, punctuated by short episodes of circulation. The second argument (CER) appears to be more stable than the first (ILR), with the latter showing evidence of slow circulation. We see no obvious correlation between the behaviour of these arguments and the times of either the closest approaches between Prometheus and Pandora (at apse anti-alignment) or the switches in the orbits of Janus and Epimetheus, suggesting that the dominant dynamical influence on Atlas is Prometheus alone. 

We investigated this further by repeating the numerical simulation for a system consisting of Atlas, Prometheus and Saturn only. In Figure 13, we show the orbital elements for Atlas derived from this three-body simulation, for comparison with those derived using the full model (Figure 4). The behaviour of the semi-major axis, mean longitude, eccentricity and longitude of pericentre are qualitatively similar, showing the same periodicity, while in Figure 14, we see that the behaviour of the resonant arguments from the three-body system are also in broad agreement with the full numerical model (compare with Figure 12). Exact agreement would not be expected, because the system is chaotic, as we demonstrate below. We conclude from this that, dynamically, Atlas is dominated by the resonant perturbation due to Prometheus alone. The secondary effect on Atlas of the known chaotic interaction between Prometheus and Pandora is much weaker, at least over the observation timespan.

\subsection{Confirmation of the chaotic orbit of Atlas}

The chaoticity of the motion of Saturn's small satellites was investigated using the Fast Lyapunov Indicator method, hereafter FLI \citep{Fro97,Fro00}. To compute the FLI time evolution, the variational equations were integrated simultaneously with the full equations of motion in a Saturn-centered cartesian reference frame, using a 
15th-order Radau type algorithm \citep{Eve85}. The initial tangent vector was chosen to be orthogonal to the force \citep{Bar09} and normalized to unity. The effects of the planet's oblateness were taken into account up to and including terms in $J_6$. State vectors were converted to geometric orbital elements using the algorithm given by \citet{Ren06}. Unlike osculating elements, the geometric elements are not contaminated by the short-period terms caused by planetary oblateness \citep{Bor94}.  We used the physical parameters of Saturn given in Table \ref{tbl2}, and masses and initial states for the satellites from Table \ref{tbl5} and Table \ref{tbl6}, respectively.

The Lyapunov characteristic exponents theory is a widely used tool for the estimation of chaoticity in dynamical systems. These exponents basically measure the rate of exponential divergence between neighboring trajectories in phase space. Chaos indicator methods based on the Lyapunov Characteristic Exponents are reviewed in \citet{Sko10}. 
Comparative studies of variational chaos indicators are given in e.g. \citet{Maf11} and \citet{Dar12}. 

For a given autonomous dynamical system $\displaystyle \dot{\bf{x}}=\bf{f}(\bf{x})$, the maximum Lyapunov characteristic exponent (hereafter MLCE) is defined by $\displaystyle \rm{MLCE}= \underset{t \rightarrow \infty}{\rm{lim}} \frac{1}{t} \ln ||\bf{w}(t)||$, where $\bf{w}$ is a deviation/tangent vector solution of the variational equations 
$\displaystyle \dot{\bf{w}} = \frac{\partial \bf{f}}{\partial \bf{x}}(\bf{x}) \bf{w} $ of the system. 
The system is then chaotic for a strictly positive value of the MLCE. The inverse of the MLCE defines a characteristic timescale, the Lyapunov time, which in practice measures the time needed for nearby orbits of the system to diverge by $e$. The FLI is also based on the computation of the norm of the tangent vector $\bf{w}$, and is defined as the initial part, up to a stopping time $t_f$, of the computation of the MLCE : $\displaystyle \rm{FLI}= \underset{0<t<t_f}{sup} \ln ||\bf{w}(t)|| $. This makes the FLI indicator a faster tool than the MLCE to distinguish chaotic from regular orbits. For a chaotic orbit, the FLI grows linearly with time, whereas the FLI of a regular orbit has a logarithmic growth. 

Typical results from numerical integrations are presented in Figures \ref{fli_prpa} and \ref{fli_atlas}. Figure \ref{fli_prpa} displays the FLI time evolution for Prometheus (solid line) and Pandora (dashed), using the physical parameters from 
Tables \ref{tbl2} and \ref{tbl5}, and the initial conditions of Table \ref{tbl6} at epoch 2007 June 1, 00:00:00.0 UTC. The upper curves represent the chaotic orbits for a three-body model consisting of Saturn, Prometheus, Pandora, and the lower curves regular orbits obtained by removing Prometheus or Pandora. The slope of the FLI evolution is in good agreement with the Lyapunov exponent value for the Prometheus-Pandora system of order $0.3$ year$^{-1}$ given by \citet{Gold03a}. In figure \ref{fli_atlas}, we show the FLI versus time for Atlas, using also the physical parameters from Tables \ref{tbl2} and \ref{tbl5}, and the initial conditions of Table \ref{tbl6} at epoch 2007 June 1, 00:00:00.0 UTC. The three upper curves represent chaotic orbits and were obtained for a model consisting of Saturn, Atlas, Prometheus, Pandora and Mimas (solid line), or Saturn, Atlas, Prometheus only (dashed and dotted). The initial semi-major axis differs within the measurement error bars by $\pm 0.1$ km. The lower solid curve is a regular orbit, obtained for a model consisting of Saturn, Atlas, Pandora and Mimas, with Prometheus removed.

We conclude therefore that Atlas's orbit is chaotic, as a direct consequence of the coupled $54:53$ resonant 
interaction with Prometheus, with a Lyapunov time of $\sim$10 years indicated by the FLI evolution. 

\section{Summary and Discussion}

Taking advantage of the extended time-span (Feb 2004 to Aug 2013) made possible by the new astrometric observations presented here, an improved orbital fit to Cassini ISS observations of the Atlas, Prometheus, Pandora, Janus and Epimetheus has been obtained, allowing their masses to be estimated with an order-of-magnitude reduction in uncertainty compared to the current published values \citep{Jac08}. 

Using the initial states and masses derived from the fit to the observations, we have then developed an improved high-precision numerical model for the orbits of these five satellites, by numerically integrating the full equations of motion in 3D between the years 2000 and 2020, incorporating perturbations from the major Saturnian satellites, Jupiter, and the Sun, together with the effects of Saturn's oblateness up to terms in J6.

Based on the new model, a numerical analysis of the key resonant interactions currently existing between these satellites has further underlined the dynamical complexity of the orbits of the small inner satellites of Saturn. 

We find that during each four-year period when Epimetheus is the innermost of the Janus-Epimetheus co-orbital pair, it is simultaneously librating in the 17:15 mean motion resonance with Prometheus, with argument $17\lambda^{\prime}-15\lambda-\varpi^{\prime}-\varpi$, and the 21:19 mean motion resonance with Pandora, with argument $21\lambda^{\prime}-19\lambda-2\varpi$. Conversely, for the intervening four-year period when Epimetheus is in the outer configuration with respect to Janus, no state of resonance exists. Although Prometheus and Pandora are chaotic, with a Lyapunov time about 3 years \citep{Gold03a,Coop04}, we were not able to detect chaotic behaviour on a similar timescale in the orbits of Janus and Epimetheus using the FLI indicator. This is probably because the relevant resonant interactions are second-order and also the masses of Janus and Epimetheus are considerably larger than those of Prometheus and Pandora (Table 6).

We also find, based on the improved numerical model presented here, that the chaotic interactions between Prometheus and Pandora studied previously by \citet{Gold03a,Gold03b,Ren03,Coop04,Ren05} are not so clearly correlated with times of anti-alignment as was apparent using the observational data available in 2004. This is also apparent from the resonant arguments: we see no obvious correlation between times of apse anti-alignment and changes from libration to circulation or vice versa. This appears to be consistent with the predictions of \citet{Farm06} who showed analytically using a pendulum approach, and numerically using the equations of motion for a 3-body system, that chaotic interactions can occur at times other than apse-alignment and that the system can drift between these regimes on a timescale of a few years.

Finally, we show that Atlas is simultaneously librating in both the 54:53 CER and 54:53 ILR with Prometheus, making it yet another example of coupled CER/ILR resonance in the Saturn system, in common with Aegaeon, Anthe and Methone. We also show that, dynamically, the behaviour of Atlas can be well-approximated by a three-body subsystem consisting of Atlas, Prometheus and Saturn only and that this system is chaotic, with a Lyapunov time of about 10 years. Future work will involve a detailed analytical study of this system.

\acknowledgments

This work was supported by the Science and Technology Facilities Council (Grant No. ST/F007566/1) and Cooper and Murray are grateful to them for financial assistance. Murray is grateful to The Leverhulme Trust for the award of a Research Fellowship. The authors thank the members and associates of the $Cassini$ ISS team, particularly Kevin Beurle for assisting in the design of the image sequences used in this work. Thanks also to Bob Jacobson for discussions. Cooper expresses gratitude to University of Lille 1 for additional funding while he was an invited researcher at the Lille Observatory. Renner thanks Marc Fouchard for discussions concerning numerical indicators of chaos. Cooper, Renner and Murray also thank their colleagues in the Encelade working group at the IMCCE of the Paris Observatory (http://www.imcce.fr/$\sim$lainey/Encelade.htm). Finally, the authors are very grateful to the anonymous referee for many useful comments and suggestions which have helped to improve this paper.

\clearpage

\begin{deluxetable}{lccccccc}
\tablecaption{Summary of Cassini ISS Observations\label{tbl1}}
\tablewidth{0pt}
\tablecolumns{6}
\tabletypesize{\scriptsize}
\rotate
\tablehead{Satellite&Start&End&Total No.\tablenotemark{a}&COL& COF&RMS\tablenotemark{b}&RMS\tablenotemark{b}\\
&&&&&&line (km)&sample (km)}
\startdata
Atlas&2004 MAY 19 15:13:24.884& 2013 AUG 28 09:44:38.943&375&299&76&12.7&10.6\\
Prometheus&2004 APR 02 10:52:24.434& 2013 AUG 28 18:01:26.753&832&166&666&16.3&13.8\\
Pandora&2004 FEB 23 21:46:25.113& 2013 AUG 28 21:00:50.685&509&166&343&23.1&21.0\\
Janus&2004 FEB 23 21:56:55.109& 2012 DEC 18 17:28:49.841&413&113&300&27.6&19.9\\
Epimetheus&2004 FEB 16 22:14:25.048& 2012 DEC 27 01:03:09.278&451&144&307&26.2&22.8\
\enddata
\tablenotetext{a}{Totals include NAC and WAC images (the overwhelming majority are NACs). WAC counts are: Atlas (0), Prometheus(1), Pandora(3), Janus(5), Epimetheus(4). COF (centre-of-figure) refers to the number of observations limb-fitted and COL (centre-of-light) to the number of observations centroided.}
\tablenotetext{b}{RMS values are post-fit O-C in km. The smaller RMS residuals for Atlas reflect the fact that most observations for this satellite used a centroiding rather than a limb-fitting technique. See Section 3.1 for further discussion.}
\end{deluxetable}

\clearpage

\begin{deluxetable}{cccccccccc}
\tablecaption{Sample of Cassini ISS Observations\label{tbl2}\tablenotemark{a}}
\tablewidth{0pt}
\tabletypesize{\scriptsize}
\rotate
\tablehead{Image ID &Time (UTC) &RA &DEC &TWIST &Line\tablenotemark{b} &Sample\tablenotemark{b} &RA &DEC &Object\\&&&(image pointing)&&(object)&(object)&(object)&(object)&Name\\&&(deg)&(deg)&(deg)&(px)&(px)&(deg)&(deg)&}
\startdata
N1463672236 &2004 MAY 19 15:13:24.884 & 36.473373 & 10.058227 &110.109989 &  856.00 &  259.00 & 36.390893 &  9.936157 &Atlas\\
N1463672551 &2004 MAY 19 15:18:39.882 & 36.326424 &  9.900412 &110.156792 &  318.00 &  540.00 & 36.386305 &  9.932481 &Atlas\\
N1463672866 &2004 MAY 19 15:23:54.880 & 36.547986 &  9.813732 &110.256168 &  842.50 &  992.00 & 36.381789 &  9.929112 &Atlas\\
N1463731456 &2004 MAY 20 07:40:24.498 & 36.431500 &  9.944629 &110.166061 &  637.00 &  274.00 & 36.418984 &  9.853239 &Atlas\\
N1463731771 &2004 MAY 20 07:45:39.496 & 36.416473 &  9.691020 &110.237587 &  326.00 &  937.00 & 36.425826 &  9.850107 &Atlas\\
N1463930793 &2004 MAY 22 15:02:40.200 & 36.471507 &  9.969432 &110.157877 &  535.00 &  485.00 & 36.467002 &  9.958112 &Atlas\\
N1463931723 &2004 MAY 22 15:18:10.194 & 36.433968 &  9.855041 &110.208023 &  383.00 &  752.00 & 36.447039 &  9.947762 &Atlas\\
N1463990463 &2004 MAY 23 07:37:09.811 & 36.493429 &  9.973714 &110.154241 &  734.00 &  257.00 & 36.451207 &  9.865373 &Atlas\\
\enddata
\tablenotetext{a}{Table 2 is published in its entirety in the electronic edition of this paper. A portion is shown here for guidance regarding its form and content. Angular coordinates are expressed with respect to the International Celestial Reference Frame (ICRF). Columns 3, 4 and 5 together provide the corrected pointing direction and orientation of the camera optical axis.}
\tablenotetext{b}{The origin of the line,sample coordinate system is at the top left of the image with line, y, increasing downwards and sample, x, to the right.}
\end{deluxetable}

\clearpage

\begin{deluxetable}{lll}
\tablecaption{Saturn constants used in orbit determination \label{tbl3}}
\tablewidth{0pt}
\tablehead{Constant&Value\tablenotemark{a}&units}
\startdata
 Pole (RA,Dec)&  (40.59550, 83.53812) &deg\\
 Pole epoch &1980 NOV 12 23:47:23.0 & \\
 Pole pression rate (RA) & - 0.04229 & deg/century\\
 Pole pression rate (DEC) & - 0.00444 & deg/century\\
 $GM$  & 3.793120706585872E+07  &km$^3$ s$^{-2}$\\
 Radius, $R_s$          &60330  &       km\\
 $J_{2}$         &1.629084747205768E-0& \\
 $J_{4}$         &$-$9.336977208718450E-04& \\
 $J_{6}$         &9.643662444877887E-05& \\
\enddata
\tablenotetext{a}{Pole position and rates from \citet{Jac04}. This position was precessed to the fit epoch, 2007 JAN 01 00:00:00.0 UTC, using the rates shown. Reference radius from \citet{Kliore80}. Zonal harmonics and $GM$ from the JPL SAT351 ephemeris.}
\end{deluxetable}

\clearpage
\begin{deluxetable}{lc}
\tablecaption{$GM$ values for other perturbing bodies used in orbit determination and numerical modelling\label{tbl4}}
\tablewidth{0pt}
\tablehead{Body& $GM$\tablenotemark{a} (km$^3$ s$^{-2}$) }
\startdata
Sun &  0.132712440040945E+12 \\
Jupiter & 0.126712764800000E+09 \\
\tableline
Mimas & 2.502784093954375E+00 \\
Enceladus & 7.211597878640501E+00  \\
Tethys & 4.121706150116760E+01 \\
Dione & 7.311724187382050E+01 \\
Rhea & 1.539395338114438E+02 \\
Titan& 8.978142565744738E+03 \\
Hyperion & 3.720643391175467E-01 \\
Iapetus & 1.205106368043288E+02 \\
\enddata
\tablenotetext{a}{GM values for the Saturnian satellites from JPL SAT351 ephemeris. GM values for Sun and Jupiter from cpck26Apr2012.tpc.}
\end{deluxetable}

\clearpage
\thispagestyle{empty} 

\begin{deluxetable}{|c|l|c|}
\tabletypesize{\scriptsize}
\tablecaption{Solutions for the planetocentric state vectors in the ICRF at epoch 2007 JUN 01 00:00:00.0 UTC  (2007 JUN 01 00:01:05.184 or  2454252.50075446 JD TDB). A machine-readable version of the table is available online. \label{tbl5}}
\tablewidth{0pt}
\tablehead{Atlas&&Units}
\startdata
$x$&±0.211131940542735E+05$\pm$?1.4635592526&km\\
$y$&±0.135369294031375E+06$\pm$?0.3073324691&km\\
$z$&$-$0.117994571363612E+05$\pm$?0.8603172593&km\\
$\dot{x}$&$-$0.163985169014096E+02$\pm$?0.0001739128&km\,s$^{-1}$\\
$\dot{y}$&±0.265851621781565E+01$\pm$?0.0000060008&km\,s$^{-1}$\\
$\dot{z}$&±0.121515798698684E+01$\pm$?0.0000933200&km\,s$^{-1}$\\
\tableline
Prometheus&&Units \\
\tableline
$x$&$-$0.139075476234052E+06$\pm$?0.6523101432&km\\
$y$&$-$0.299630283368139E+04$\pm$?0.1379426749&km\\
$z$&±0.121750245927928E+05$\pm$?0.5254326536&km\\
$\dot{x}$&±0.437822964122312E+00$\pm$?0.0000727584&km\,s$^{-1}$\\
$\dot{y}$&$-$0.164562743532697E+02$\pm$?0.0000041728&km\,s$^{-1}$\\
$\dot{z}$&±0.117690166176359E+01$\pm$?0.0000627568&km\,s$^{-1}$\\
\tableline
Pandora&&Units \\
\tableline
$x$&±0.882713818598140E+05$\pm$?1.0113932949&km\\
$y$&±0.110230473398456E+06$\pm$?0.8085330496&km\\
$z$&$-$0.158256977509867E+05$\pm$?0.7225425073&km\\
$\dot{x}$&$-$0.127680928599784E+02$\pm$?0.0001101016&km\,s$^{-1}$\\
$\dot{y}$&±0.102096992841201E+02$\pm$?0.0000603615&km\,s$^{-1}$\\
$\dot{z}$&±0.353489290599632E+00$\pm$?0.0000797166&km\,s$^{-1}$\\
\tableline
Janus&&Units \\
\tableline
$x$&±0.117969400691064E+06$\pm$?1.1338663430&km\\
$y$&±0.925426967020695E+05$\pm$?0.7704515369&km\\
$z$&$-$0.173829056177225E+05$\pm$?0.6284654451&km\\
$\dot{x}$&$-$0.990085822183523E+01$\pm$?0.0001120748&km\,s$^{-1}$\\
$\dot{y}$&±0.124533083800756E+02$\pm$?0.0000386836&km\,s$^{-1}$\\
$\dot{z}$&$-$0.803656578358153E-01$\pm$?0.0000722691&km\,s$^{-1}$\\
\tableline
Epimetheus&&Units \\
\tableline
$x$&±0.188797545005757E+05$\pm$?1.6617867438&km\\
$y$&$-$0.149365181703202E+06$\pm$?0.1356577159&km\\
$z$&±0.103158087510301E+05$\pm$?0.8851737386&km\\
$\dot{x}$&±0.157497527566054E+02$\pm$?0.0001515017&km\,s$^{-1}$\\
$\dot{y}$&±0.174499934532077E+01$\pm$?0.0000416942&km\,s$^{-1}$\\
$\dot{z}$&$-$0.148701850883106E+01$\pm$?0.0000860619&km\,s$^{-1}$\\
\tableline
rms& 0.855 (NAC)&pixel\\
rms& 1.057 (NAC)&arcsec\\
\tableline
\enddata
\end{deluxetable}

\clearpage

\begin{deluxetable}{lcc}
\tabletypesize{\small}
\tablewidth{0pt}
\tablecaption{Solutions for $GM$ (km$^3$ s$^{-2}$ $\times$ 10$^{3}$)\label{tbl6}}
\tablehead{Satellite  & Jacobson et al (2008) & This work}
\startdata
Atlas & 0.44 $\pm$0.04 & 0.384 $\pm$0.001 \\
Prometheus & 10.64 $\pm$0.10 & 10.677 $\pm$0.006 \\
Pandora & 9.15 $\pm$0.13 & 9.133 $\pm$0.009 \\
Janus & 126.60 $\pm$0.08 & 126.51 $\pm$0.03 \\
Epimetheus & 35.13 $\pm$0.02 & 35.110 $\pm$0.009 \\
\enddata
\end{deluxetable}

\clearpage

\setlength{\topmargin}{0in}

\setcounter{figure}{0}

\begin{figure}
\epsscale{0.8}
\plotone{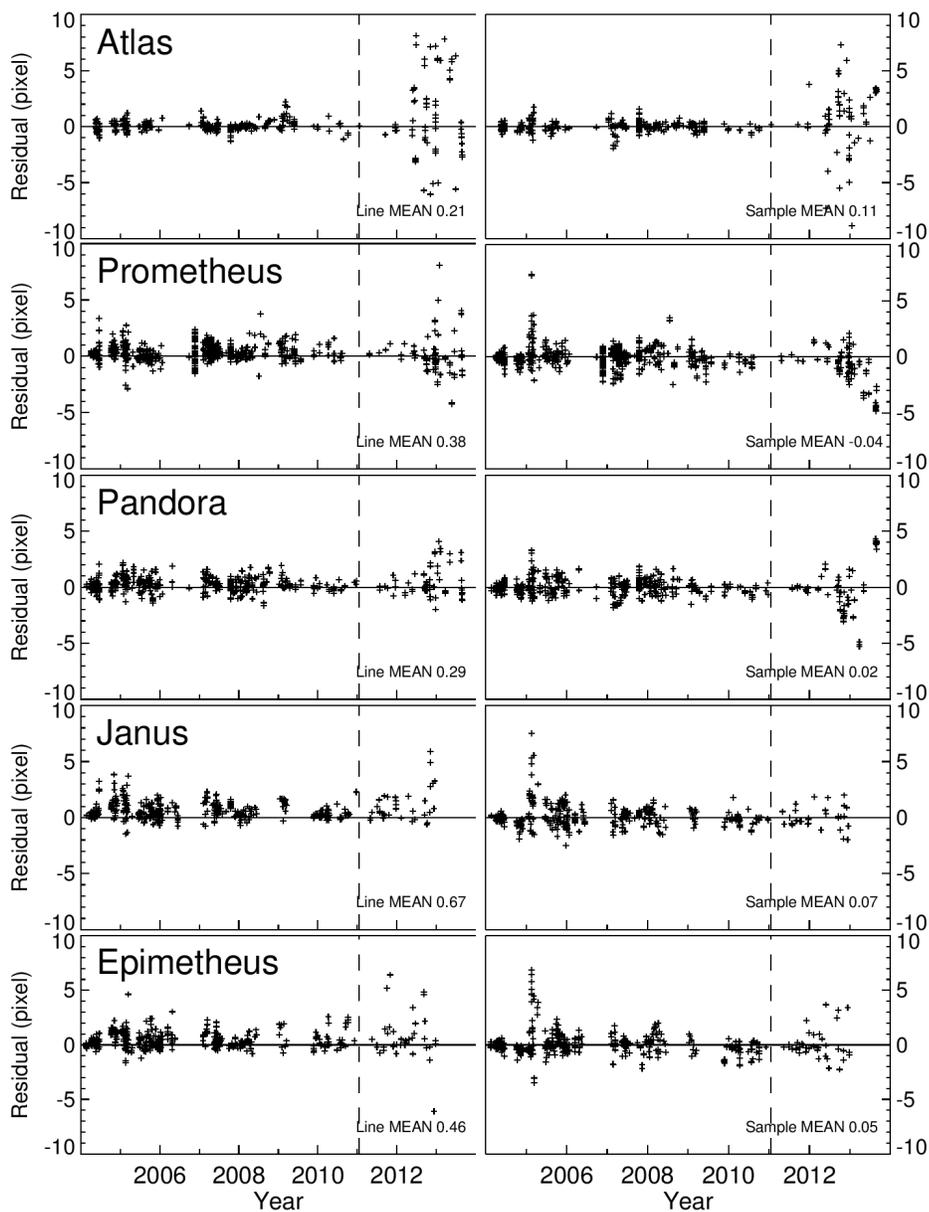}
\label{Fig 1}
\caption{Pre-fit residuals (O-C). Line and sample residuals relative to JPL's SAT353 ephemeris, plotted in NAC pixels as a function of time. For each satellite, the line residuals are shown in the left-hand panel and sample residuals in the right. Vertical and horizontal scales match those in Figure 2. Vertical dashed lines correspond to 2011 JAN 20, the end of the observation timespan used in the SAT353 solution. The divergence in the residuals beyond this time is due to the lack of observational constraint (SAT363 shows no divergence because it is a fit to observations up to late 2013.)}
\end{figure}

\clearpage

\begin{figure}
\epsscale{0.8}
\plotone{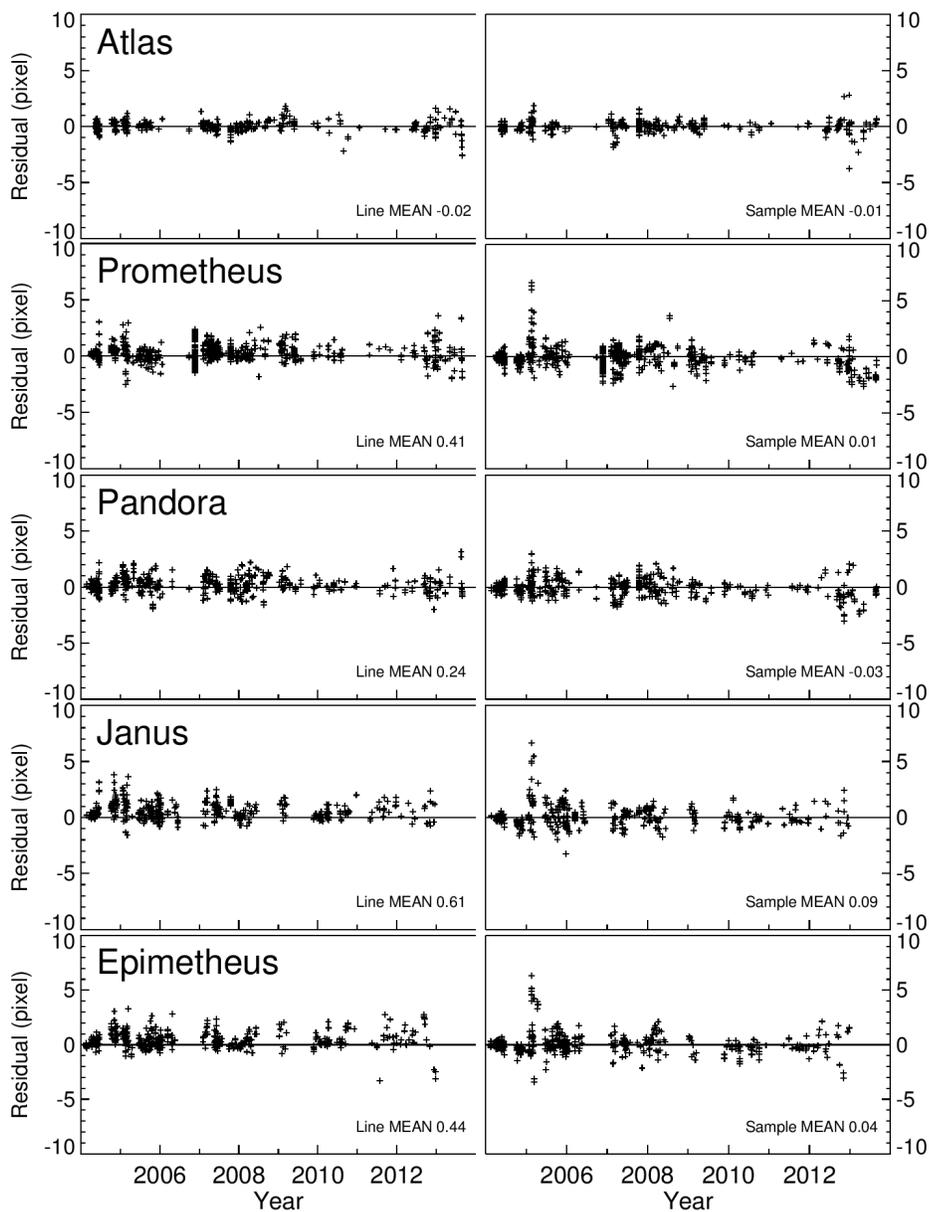}
\label{Fig 2}
\caption{Post-fit residuals (O-C). Line and sample residuals after fitting to the new observations, plotted in NAC pixels as a function of time. The divergence visible in the pre-fit residuals (Figure 1) is now greatly reduced due to the better observational coverage. For each satellite, the line residuals are shown in the left-hand panel and sample residuals in the right. Vertical and horizontal scales match those in Figure 1.}
\end{figure}

\clearpage

\begin{figure}
\epsscale{0.7}
\plotone{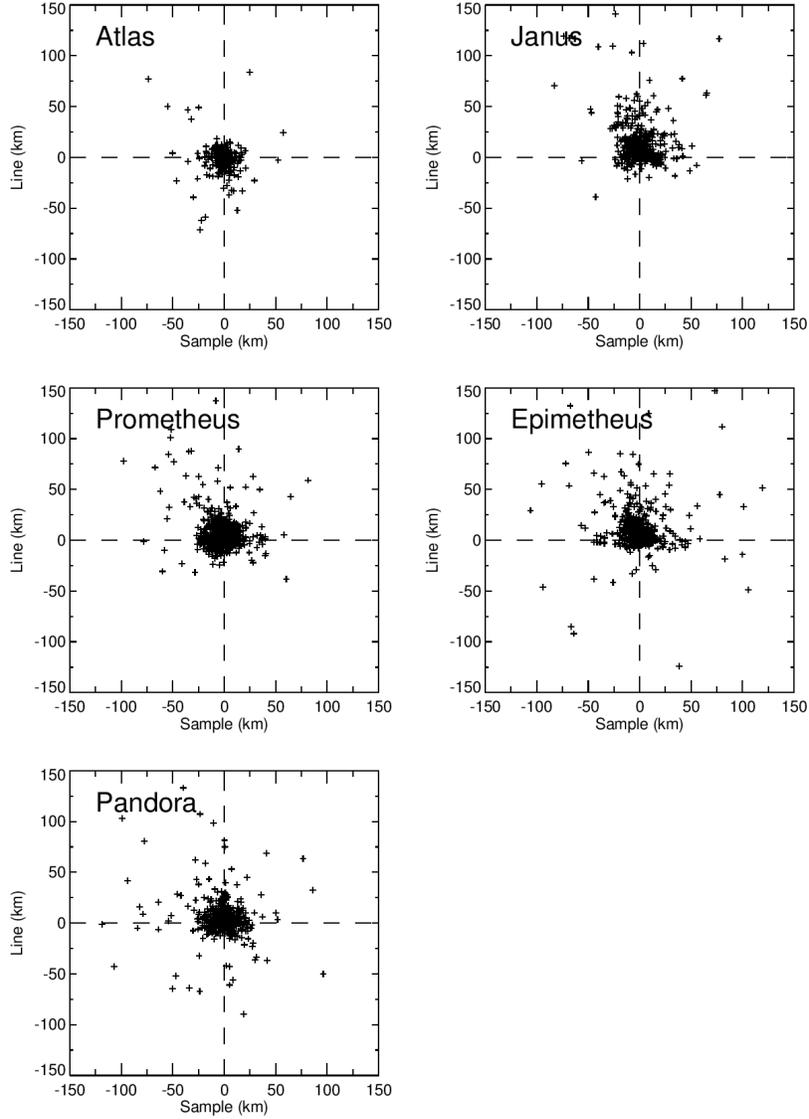}
\label{Fig 3}
\caption{Scatter plot showing post-fit residuals (O-C) in km. Data are as Figure 2, except plotted as line residual versus sample residual, converted to km. Four line values and two sample values are off-scale. These correspond to low resolution images (including some WAC images), where the number of km/pixel is particularly large. The small bias in the line direction, particularly for Janus and Epimetheus, arises in the limb-detection algorithm used to measure centres-of-figure. Atlas shows very little bias because the observations are generated mainly using a centroiding technique rather than limb-fitting. See Section 3.1 for further discussion.}
\end{figure}

\begin{figure}
\plotone{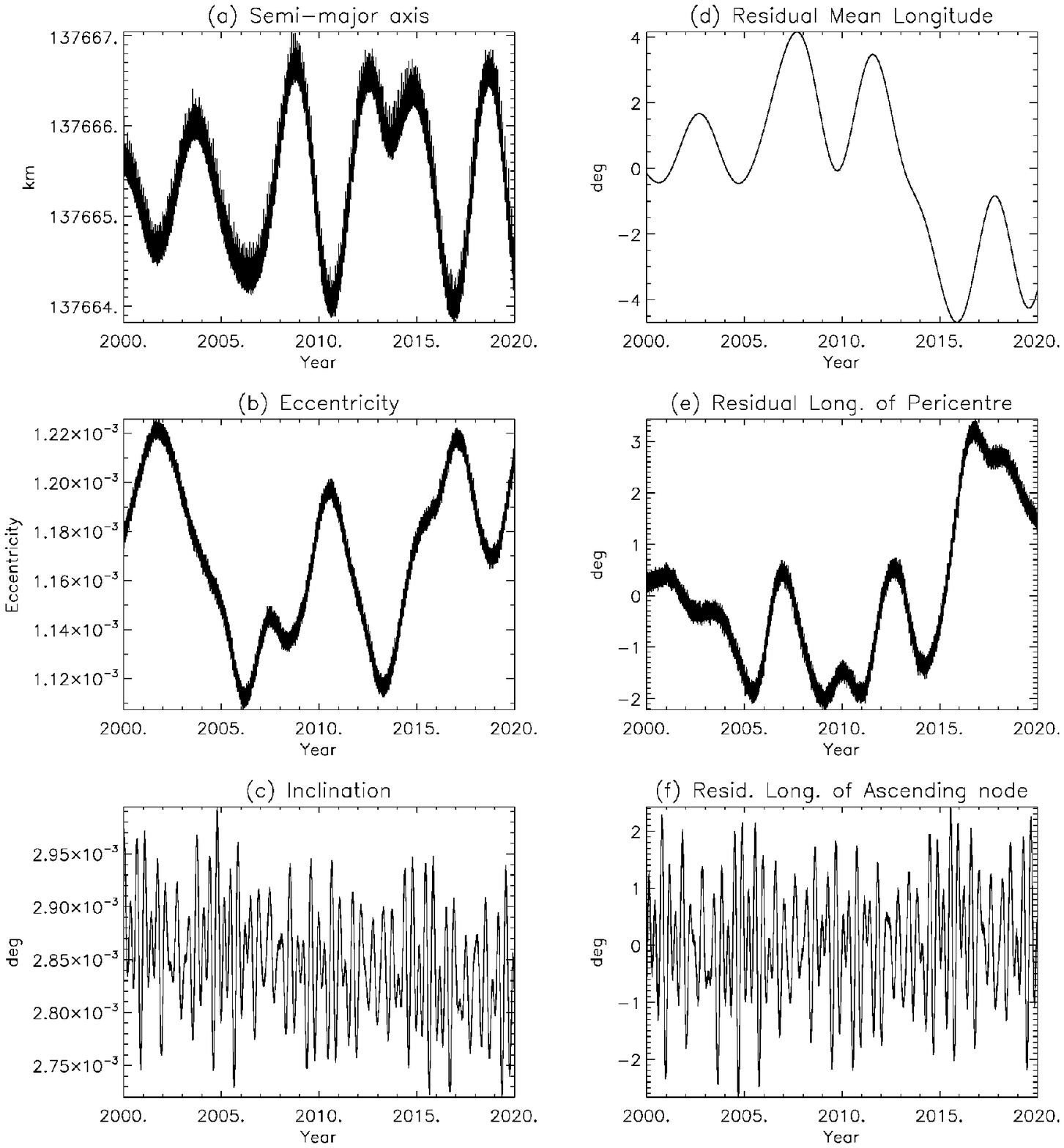}
\label{Fig 4}
\caption{Geometric orbital elements for Atlas as a function of time between 2000 JAN 01 and 2020 JAN 01, from the full numerical model, using the parameters from Tables 2 to 6 inclusive. Linear background trends have been subtracted from the mean longitude, longitude of pericenter and longitude of ascending node, using rates of 598.31312, 2.8799044 and -2.8679477 deg/day, respectively.}
\end{figure}

\begin{figure}
\plotone{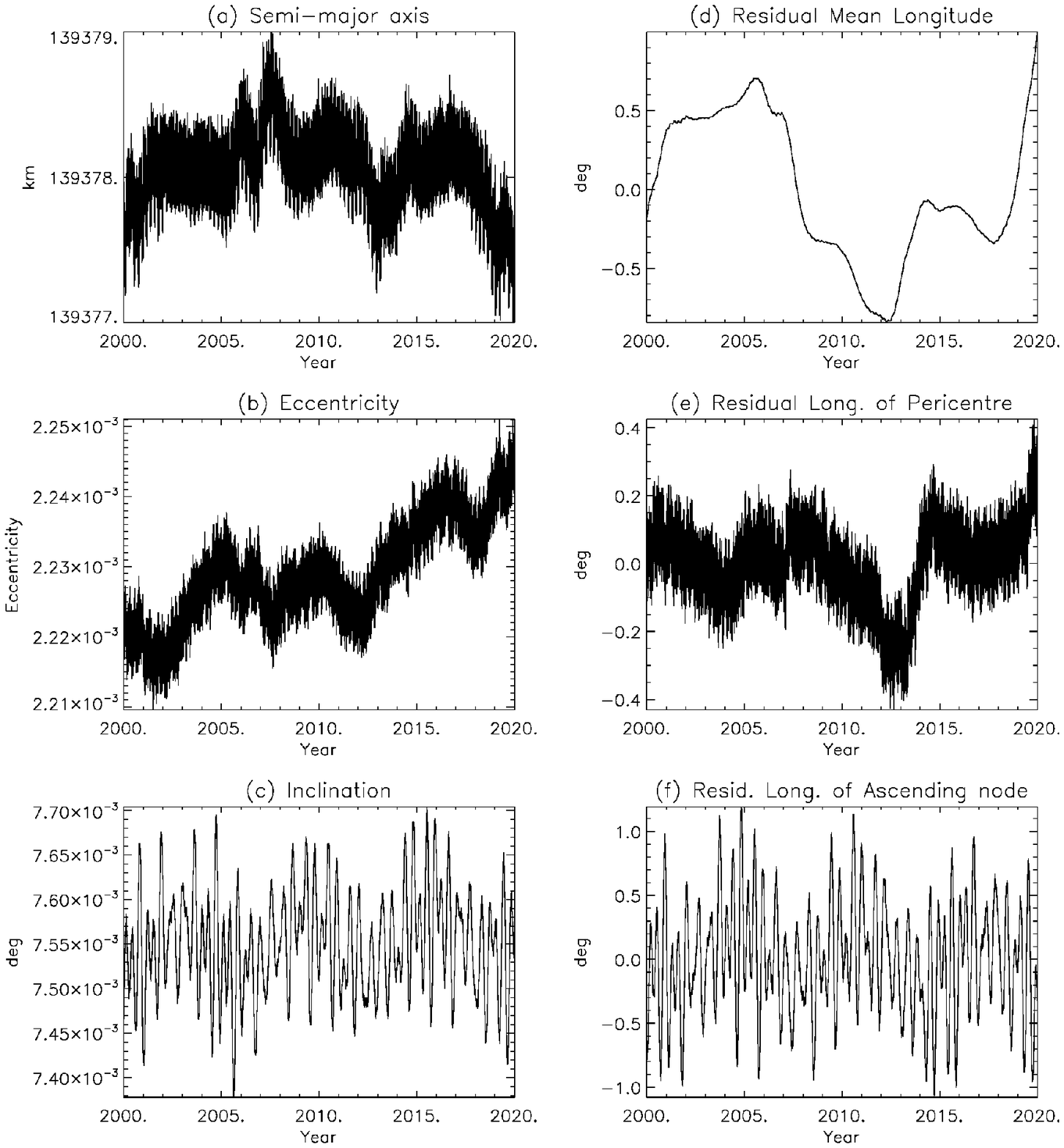}
\label{Fig 5}
\caption{Geometric orbital elements for Prometheus as a function of time between 2000 JAN 01 and 2020 JAN 01, from the full numerical model, using the parameters from Tables 2 to 6 inclusive. Linear background trends have been subtracted from the mean longitude, longitude of pericenter and longitude of ascending node, using rates of 587.28501, 2.7576978 and -2.7449045 deg/day, respectively.}
\end{figure}

\begin{figure}
\plotone{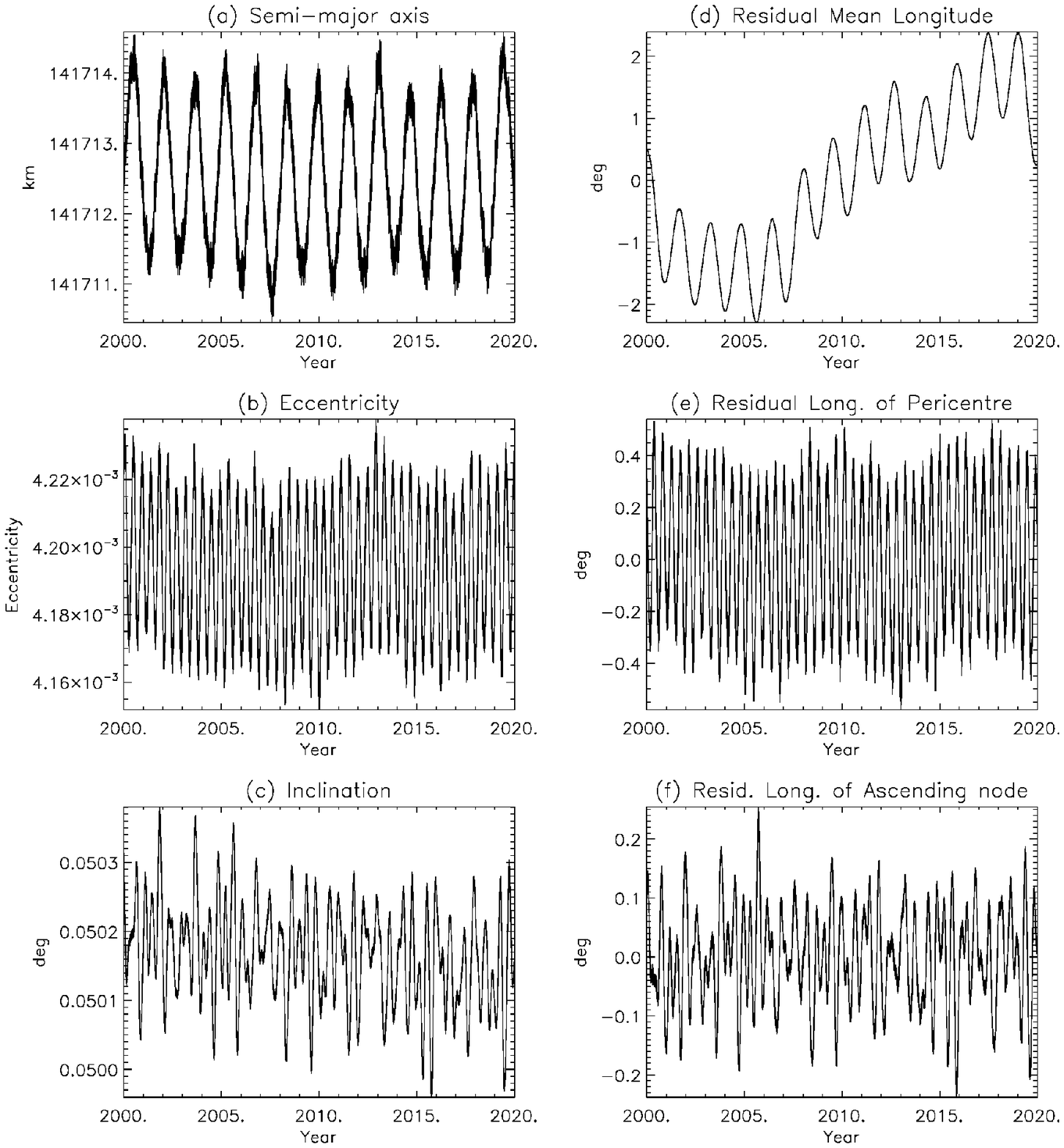}
\label{Fig 6}
\caption{Geometric orbital elements for Pandora as a function of time between 2000 JAN 01 and 2020 JAN 01, from the full numerical model, using the parameters from Tables 2 to 6 inclusive. Linear background trends have been subtracted from the mean longitude, longitude of pericenter and longitude of ascending node, using rates of 572.78861, 2.5997363and -2.5880595 deg/day, respectively.}
\end{figure}

\begin{figure}
\plotone{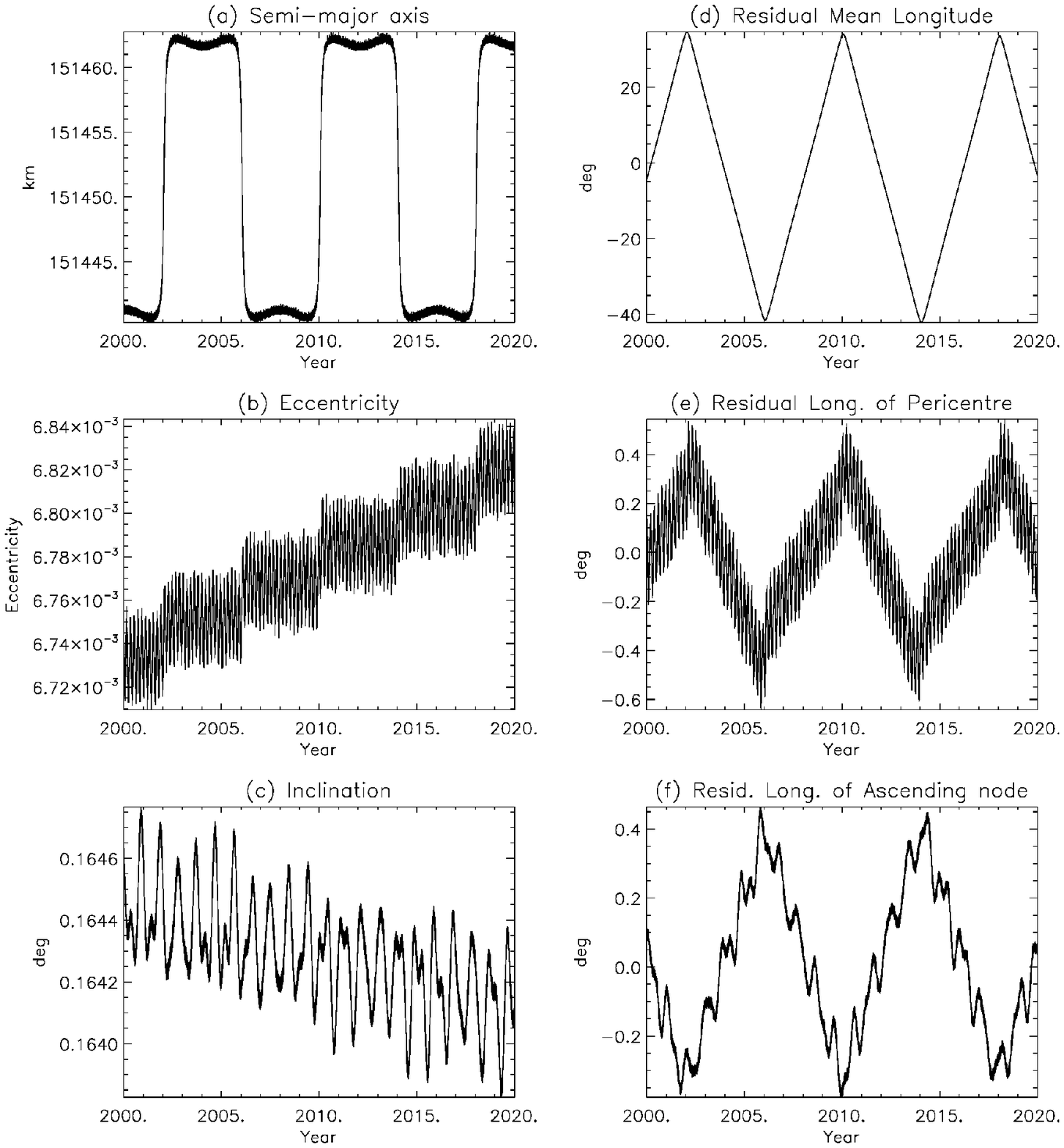}
\label{Fig 7}
\caption{Geometric orbital elements for Janus as a function of time between 2000 JAN 01 and 2020 JAN 01, from the full numerical model, using the parameters from Tables 2 to 6 inclusive. Linear background trends have been subtracted from the mean longitude, longitude of pericenter and longitude of ascending node, using rates of 518.29207, 2.0535596 and -2.0453735 deg/day, respectively.}
\end{figure}

\begin{figure}
\plotone{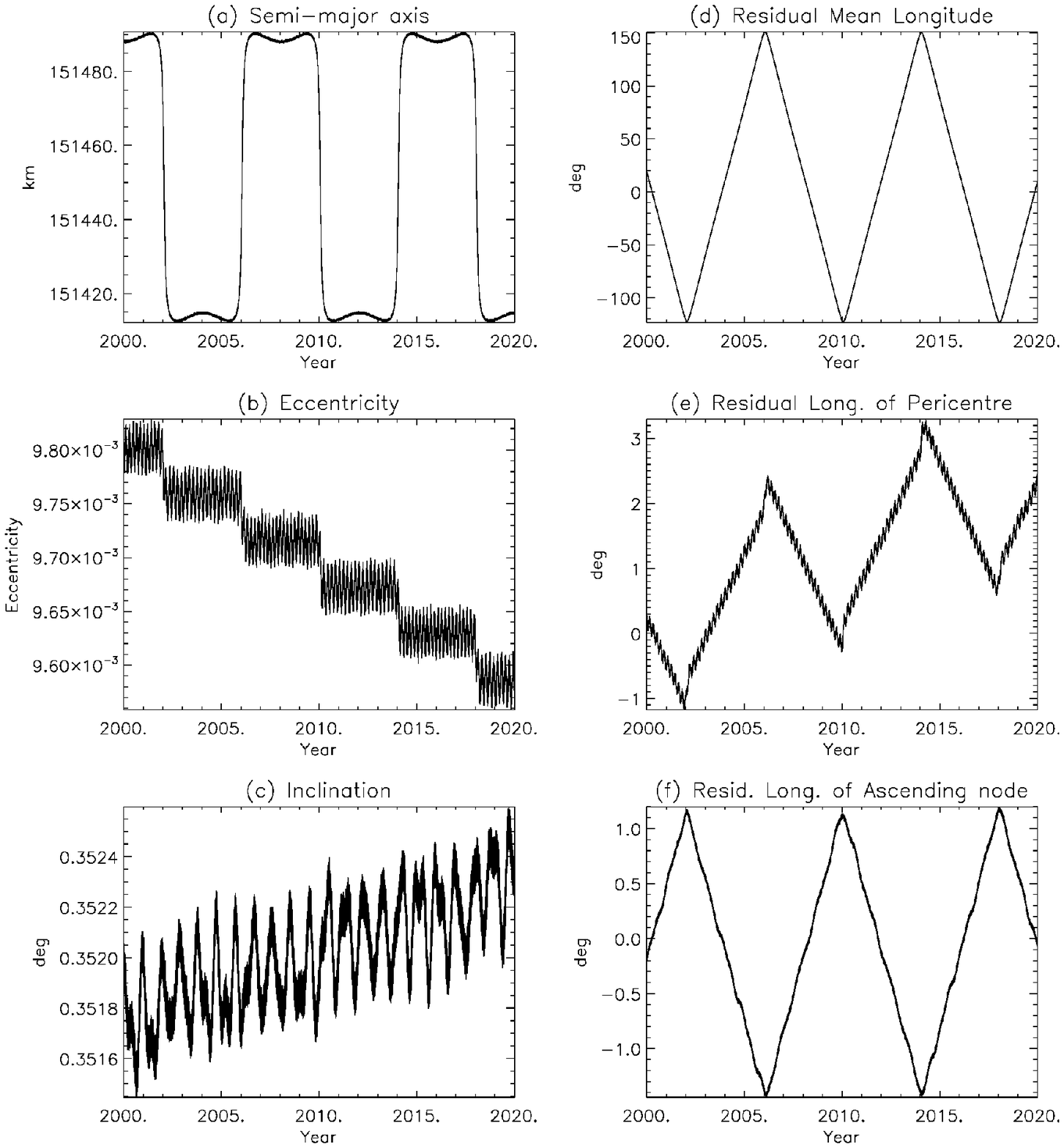}
\label{Fig 8}
\caption{Geometric orbital elements for Epimetheus as a function of time between 2000 JAN 01 and 2020 JAN 01, from the full numerical model, using the parameters from Tables 2 to 6 inclusive. Linear background trends have been subtracted from the mean longitude, longitude of pericenter and longitude of ascending node, using rates of 518.29207, 2.0538720 and -2.0455737 deg/day, respectively.}:w

\end{figure}

\begin{figure}
\epsscale{0.6}
\plotone{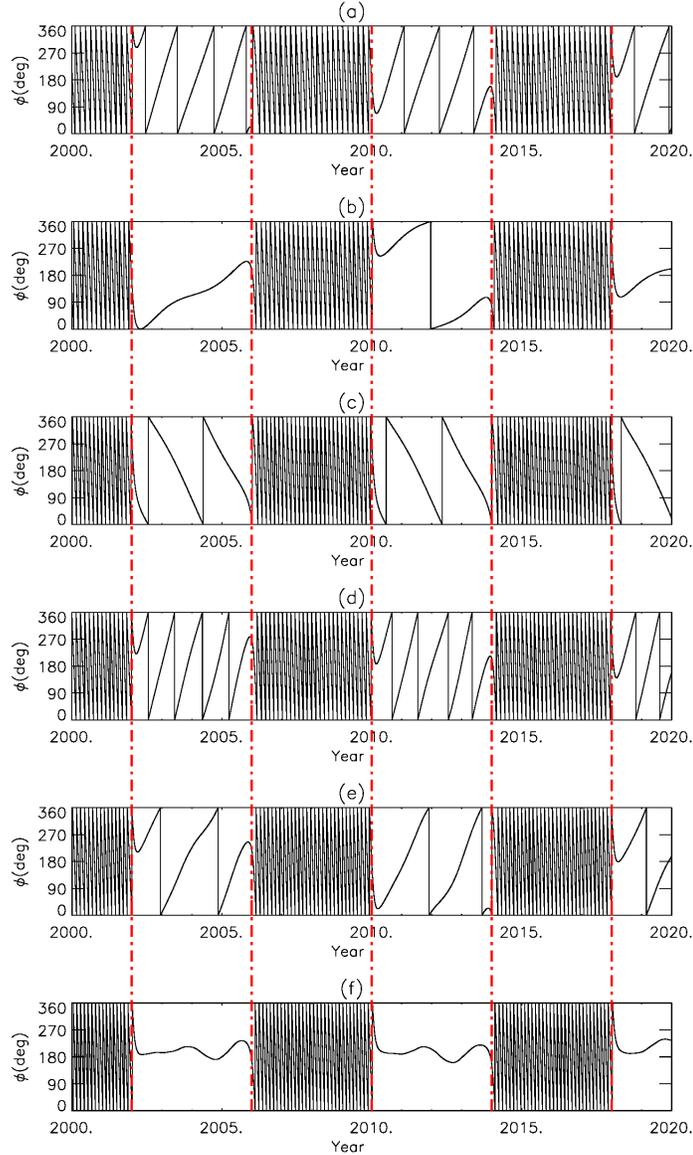}
\label{Fig 9}
\caption{Resonant arguments, $\phi$ in degrees, for Epimetheus, Prometheus and Pandora between 2000 JAN 01 and 2020 JAN 01 from the full numerical model, using the physical parameters from Tables 2 to 6 inclusive:  (a) $ \phi=17\lambda^{\prime}-15\lambda-2\varpi^{\prime} $, (b) $ \phi=17\lambda^{\prime}-15\lambda-\varpi^{\prime}-\varpi $, (c) $ \phi=17\lambda^{\prime}-15\lambda-2\varpi$ , (d) $\phi=21\lambda^{\prime}-19\lambda-2\varpi^{\prime}$,  (e) $ \phi=21\lambda^{\prime}-19\lambda-\varpi^{\prime}-\varpi$, (f) $ \phi=21\lambda^{\prime}-19\lambda-2\varpi$, where the primed quantities refer to Epimetheus in all cases, and the unprimed quantities to Prometheus for (a) to (c), and Pandora for (d) to (f). Vertical red dot-dashed lines mark switches in the configuration of Janus and Epimetheus.}
\end{figure}

\begin{figure}
\epsscale{0.6}
\plotone{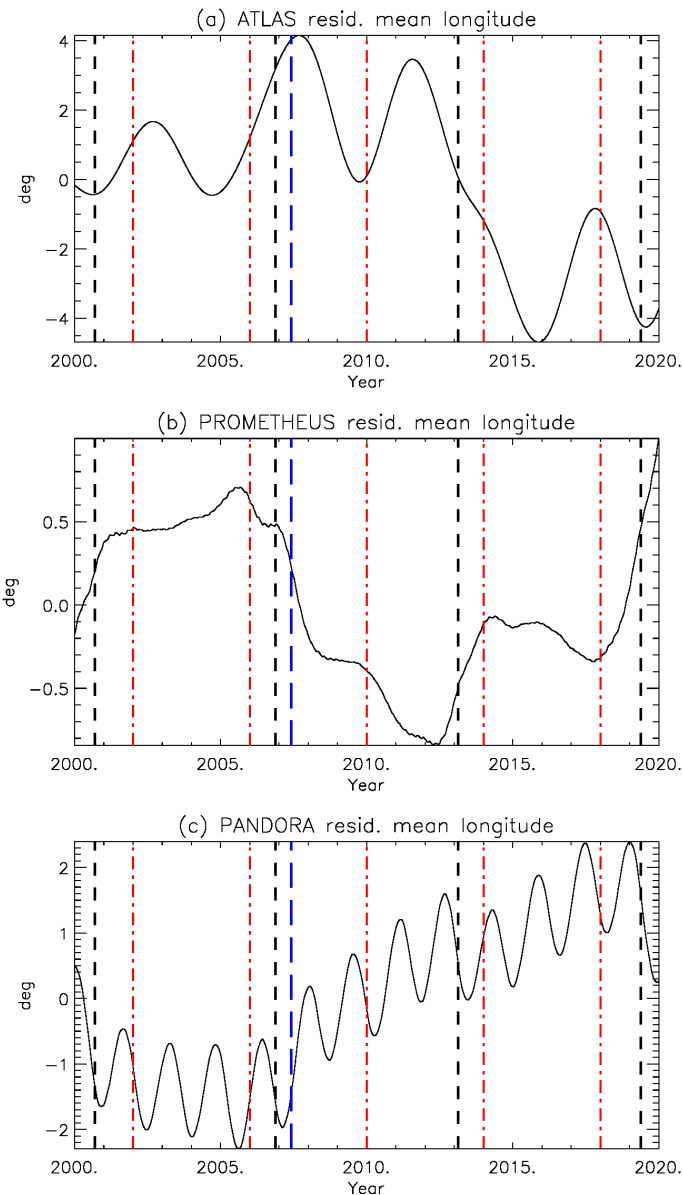}
\label{Fig 10}
\caption{Mean Longitudes for (a) Atlas, (b) Prometheus and (c)  Pandora between 2000 JAN 01 and 2020 JAN 01 from the full numerical model, using the physical parameters from Tables 2 to 6 inclusive. Plotted values are the same as Figures 4(d), 5(d) and 6(d), respectively (see captions for those figures). Vertical black dashed lines mark times of closest approach between Prometheus and Pandora. Vertical red dot-dashed lines mark times of switches in the configuration of Janus and Epimetheus. Vertical blue line marks fit epoch.}
\end{figure}

\begin{figure}
\plotone{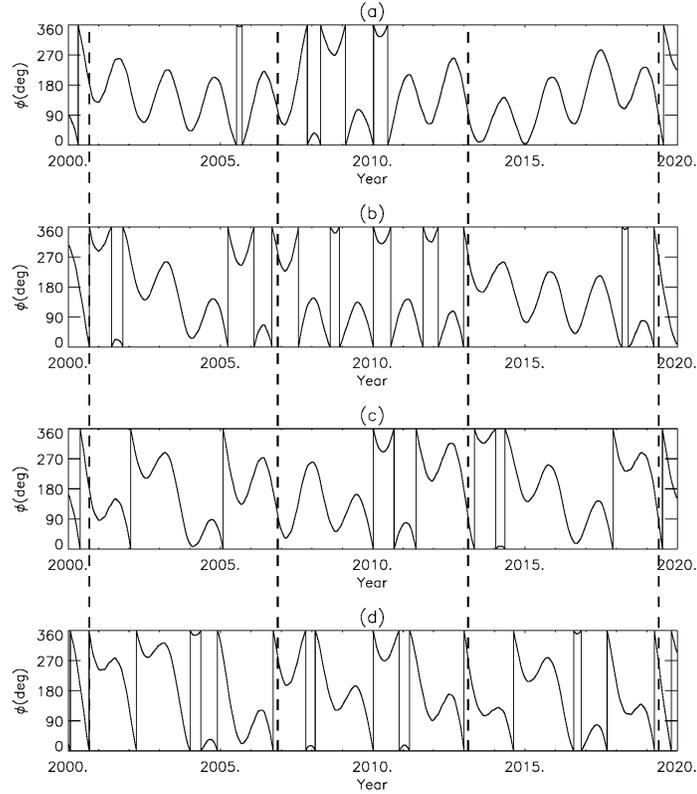}
\label{Fig 11}
\caption{Resonant arguments, $\phi$ in degrees, for Prometheus and Pandora, between 2000 JAN 01 and 2020 JAN 01 from the full numerical model, using the physical parameters from Tables 2 to 6 inclusive: (a) $ \phi=121\lambda^{\prime}-118\lambda-3\varpi^{\prime} $, (b) $ \phi=121\lambda^{\prime}-118\lambda-2\varpi^{\prime}-\varpi $, (c) $ \phi=121\lambda^{\prime}-118\lambda-\varpi^{\prime}-2\varpi$ and (d) $ \phi=121\lambda^{\prime}-118\lambda-3\varpi $, where the primed quantities refer to Pandora and the unprimed quantities to Prometheus. Vertical black dashed lines mark times of closest approach between Prometheus and Pandora.}
\end{figure}

\begin{figure}
\plotone{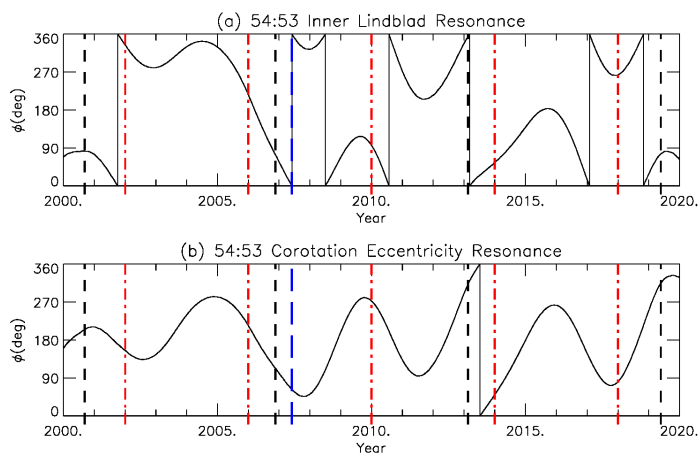}
\label{Fig 12}
\caption{Resonant arguments for Atlas and Prometheus between 2000 JAN 01 and 2020 JAN 01 from the full numerical model, using the physical parameters from Tables 2 to 6 inclusive. Resonant arguments are (a) $ \phi_1=54\lambda^{\prime}-53\lambda-\varpi $ and (b) $ \phi_2=54\lambda^{\prime}-53\lambda-\varpi^{\prime} $, where the primed quantities refer to Prometheus and the unprimed quantities to Atlas, with $\lambda$ representing the mean longitude, and $\varpi$ the longitude of pericentre in each case. Vertical black dashed lines mark times of closest approach between Prometheus and Pandora. Vertical red dot-dashed lines mark times of switches in the configuration of Janus and Epimetheus. Vertical blue line marks fit epoch.}
\end{figure}

\begin{figure}
\plotone{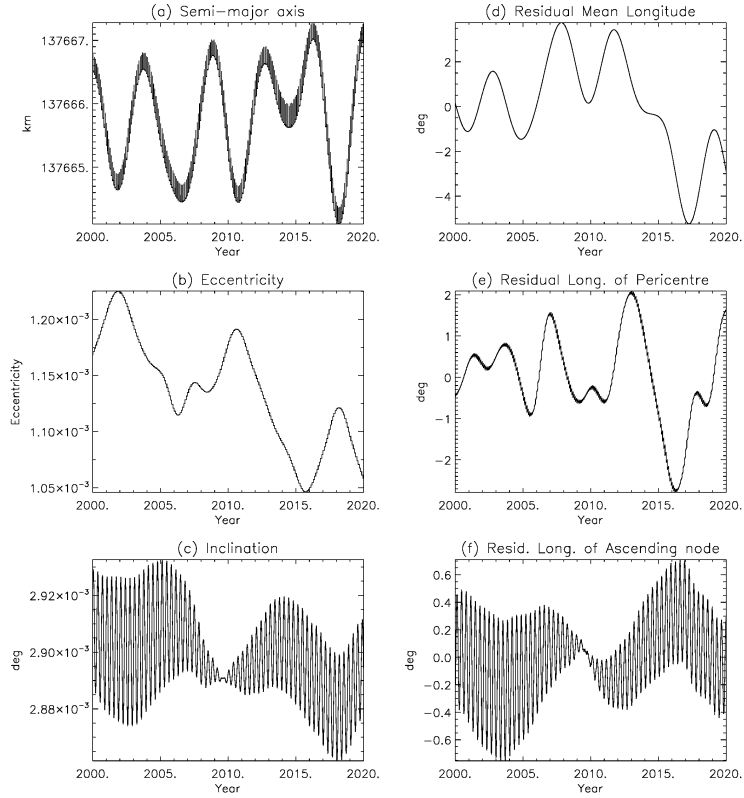}
\label{Fig 13}
\caption{Geometric orbital elements for Atlas as a function of time between 2000 JAN 01 and 2020 JAN 01 from a three-body model consisting of Atlas, Prometheus and Saturn, using selected parameters from Tables 2 to 6 inclusive. Linear background trends have been subtracted from the mean longitude, longitude of pericenter and longitude of ascending node, using rates of 598.31185, 2.8789092 and -2.8674690 deg/day, respectively.}
\end{figure}

\begin{figure}
\plotone{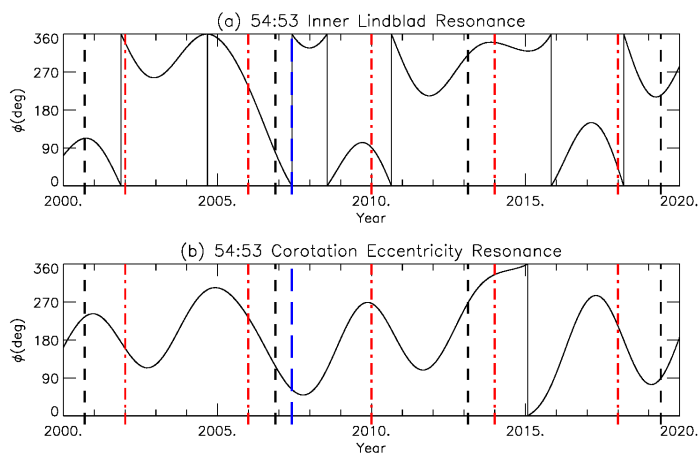}
\label{Fig 14}
\caption{Resonant arguments for Atlas and Prometheus between 2000 JAN 01 and 2020 JAN 01 from a three-body model consisting of Atlas, Prometheus and Saturn, using selected parameters from Tables 2 to 6 inclusive. Resonant arguments are (a) $ \phi_1=54\lambda^{\prime}-53\lambda-\varpi $ and (b) $ \phi_2=54\lambda^{\prime}-53\lambda-\varpi^{\prime} $, where the primed quantities refer to Prometheus and the unprimed quantities to Atlas, with $\lambda$ representing the mean longitude, and $\varpi$ the longitude of pericentre in each case. Vertical black dashed lines mark times of closest approach between Prometheus and Pandora. Vertical red dot-dashed lines mark times of switches in the configuration of Janus and Epimetheus. Vertical blue line marks fit epoch.}
\end{figure}

\begin{figure}
\centering
\includegraphics[width=.6\textwidth,angle=90]{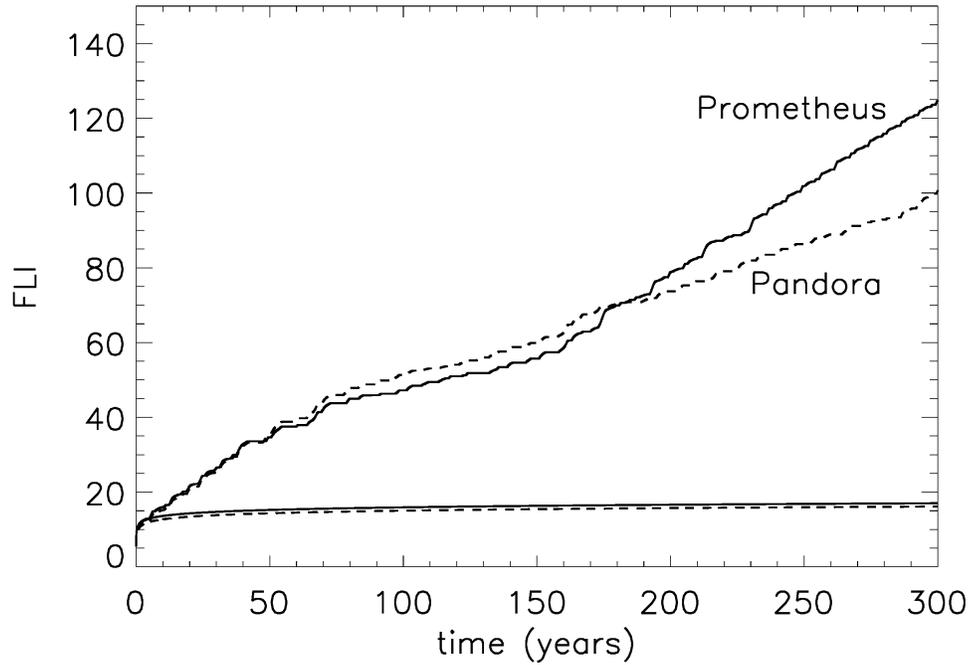}
\caption{FLI versus time for Prometheus (solid line) and Pandora (dashed), using the physical parameters from Tables \ref{tbl2} and \ref{tbl6} at epoch 2007 June 1, 00:00:00.0 UTC. The upper curves represent the chaotic orbits from a three-body model consisting of Saturn, Prometheus, Pandora, and the lower curves show regular orbits obtained by removing Prometheus or Pandora.}
\label{fli_prpa}
\end{figure}

\begin{figure}
\centering
\includegraphics[width=.6\textwidth,angle=90]{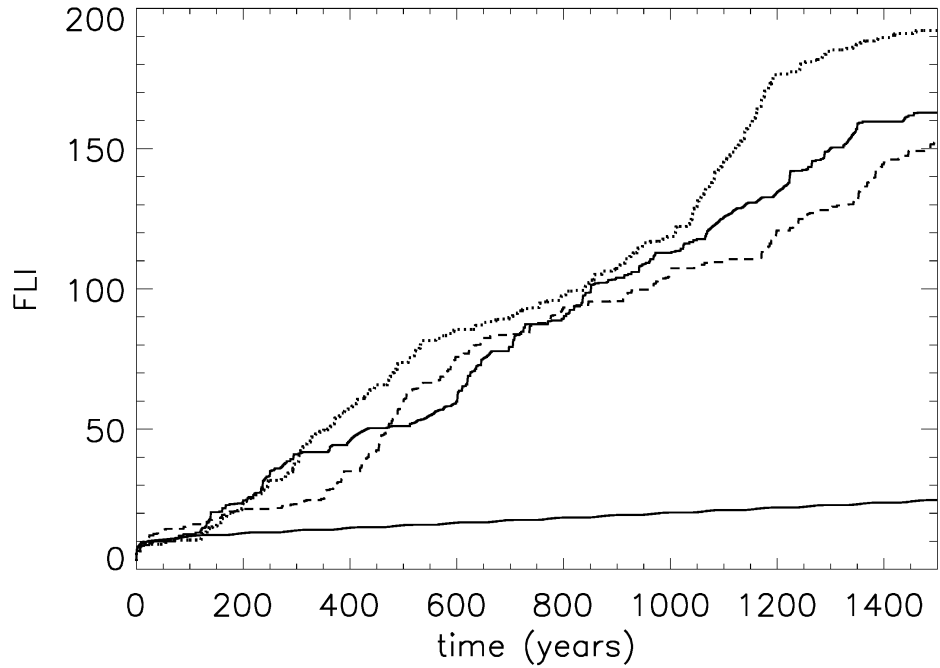}
\caption{FLI versus time for Atlas, using the physical parameters from Tables \ref{tbl2} and \ref{tbl6} at epoch 2007 June 1, 00:00:00.0 UTC. The three upper curves represent chaotic orbits and were obtained for a model consisting of Saturn, Atlas, Prometheus, Pandora and Mimas (solid line), or Saturn, Atlas, Prometheus only (dashed and dotted). The initial semi-major axis differs within the measurement error bars by $\pm 0.1$ km. The lower solid curve shows a regular behaviour, obtained for a model consisting of  Saturn, Atlas, Pandora and Mimas, with Prometheus removed.}
\label{fli_atlas}
\end{figure}

\begin{deluxetable}{l}
\tablecaption{SPICE kernels used in orbit determination and numerical modelling \label{tbl7}. Only the first 10 lines of the table are reproduced here. A machine-readable version of the full table is available in the online edition.}
\tablewidth{0pt}
\tabletypesize{\scriptsize} 
\tablehead{Kernel name\tablenotemark{a}\\}
\startdata
cas\_v40.tf\\
cas\_rocks\_v18.tf\\         
pck00007.tpc\\               
naif0010.tls\\
casT91.tsc\\
cpck26Apr2012.tpc\\
corrected\_120217\_cpck\_rock\_21Jan2011\_merged.tpc\\
de414.bsp\\
sat242.bsp\\
jup263.bsp\\
\enddata
\tablenotetext{a}{Kernels are available by anonymous ftp from ftp://naif.jpl.nasa.gov/pub/naif/CASSINI/kernels}
\end{deluxetable}

\end{document}